\documentclass[preprint,showpacs,amsmath,amssymb,prc]{revtex4}

\usepackage{epsfig}
\usepackage{psbox}
\usepackage{graphicx}

\newcommand{\smfrac}[2]{\mbox{$\frac{#1}{#2}$}}
\begin{document}

\title{Relevance of pseudospin symmetry in 
proton-nucleus scattering
}
\author{H. Leeb}
\affiliation{Atominstitut der \"Osterreichischen 
       Universit\"aten, Technische Universit\"at Wien,
        Wiedner Hauptstra\ss e 8-10, A-1040 Wien, Austria}
\author{S.\,A. Sofianos}
\affiliation{Physics Department, University of South Africa,
         P.O.Box 392, Pretoria 0003, South Africa}

\date{\today}
%
\begin{abstract}
The manifestation of pseudospin-symmetry  in proton-nucleus scattering
is discussed. Constraints on the pseudospin-symmetry violating
scattering amplitude are given which require as input cross section 
and polarization data, but no measurements of the spin rotation function. 
Application of these constraints to p\,-$^{58}$Ni and p\,-$^{208}$Pb 
scattering data in the laboratory energy range of  200\,MeV to 800\,MeV, 
reveals  a significant violation of the symmetry at lower energies
and a weak one at higher energies. Using a schematic model within the 
Dirac phenomenology, the role of the Coulomb potential in
proton-nucleus scattering with regard to pseudospin  symmetry is studied. 
Our results indicate that the existence of pseudospin-symmetry 
in proton-nucleus scattering is questionable in the whole energy region
considered and that the violation of this symmetry stems from the
long range nature of the  Coulomb interaction.
\end{abstract}
\maketitle
%
%
\section{Introduction}
Originally the concept of pseudospin was introduced on an
observational basis \cite{Hecht69,Arima69} to explain the
quasi-degeneracy of spherical shell orbitals with non relativistic 
quantum numbers ($n_r,\ell ,j=\ell+\frac{1}{2}$) and 
($n_r-1,\ell +2 ,j=\ell +\frac{3}{2}$), where $n_r$, $\ell $ and $j$ 
are the single-nucleon radial, orbital angular momentum, 
and total angular momentum quantum numbers, respectively.
This symmetry approximately persists also for deformed nuclei 
\cite{Ratna73,Draayer84,Zeng91} and even for the case of triaxiality 
\cite{Blokhin97,Beuschel97}. The origin of pseudospin-symmetry was not
well understood for a long time. Only in the nineties its
relation to the invariance of the Dirac Hamiltonian with $V_V=-V_S$
under specific $SU(2)$ transformations \cite{Bell75} has been pointed 
out
\cite{Ginocchio97,Ginocchio98a,Ginocchio98b,Lalazissis98,Ginocchio99a}.
Here, $V_S$ and $V_V$ are the scalar and vector potentials,
respectively. In the non relativistic limit this leads to a Hamiltonian
which conserves pseudospin, 
\begin{equation}
     {\bf \tilde s}=2 \frac{\bf s\cdot p}{p^2}{\bf p} - {\bf s} \, ,
\label{eq:stilde}
\end{equation}
where ${\bf s}$ is the spin and ${\bf p}$ is the momentum operator of 
the nucleon. 

A Dirac Hamiltonian with $V_V=-V_S$ does not sustain any Dirac valence 
bound state \cite{Ginocchio97}. Therefore, realistic mean fields used in
nuclear structure physics must exhibit at least a weak
pseudospin-symmetry violation. Actually the violation is smaller than
anticipated from realistic relativistic mean field calculations
\cite{Ginocchio98a}.

The question then arises whether this symmetry, associated with $V_V=-V_S$, 
manifests itself also in proton-nucleus scattering. From studies
within Dirac phenomenology, the proton-nucleus scattering is
described quite well by complex Dirac potentials with $V_V \approx -V_S$ 
\cite{Geramb85}. In 1988 Bowlin {\em et al.} \cite{Bowlin88} evaluated the 
analyzing power $P(\theta )$ and the spin rotation function 
$Q(\theta )$ under the assumption that $V_V=-V_S$, where $\theta $ is the 
scattering angle. They found a significant deviation of the 
experimental polarization and spin-rotation data from the predicted 
ones. Based on an algebraic estimate, they concluded that the symmetry 
is destroyed for low-energy proton scattering and that at high energies 
only some remnants might survive.

 Recently, Ginocchio \cite{Ginocchio99b} 
revisited this question and evaluated, in a first order approximation,
 the ratio $R_{\rm ps}$ between the pseudospin symmetry breaking and the 
non-breaking part of the scattering amplitude. By considering 
experimental p\,-$^{208}$Pb elastic scattering data at
$E_{\rm Lab}=800$\,MeV \cite{Fergerson86},  he obtained a 
relatively small pseudospin dependent 
part of the scattering amplitude at all scattering angles $\theta $
at which data were available ($\theta < 18^\circ$). This result, 
confirmed also by exact calculations \cite{Leeb00}, has 
been interpreted as an indication for the relevance of 
pseudospin-symmetry for proton-nucleus scattering -- at least at medium 
energies. However, to make more conclusive statements about this point
systematic studies of proton-nucleus scattering data covering a range 
of nuclei and energies are required. At present, such studies 
are hampered by the limited availability of complete data sets.
This is due  to the fact that  measurements of the spin rotation function 
are difficult to obtain and therefore data are and will be very scarce. 

In the present work we investigate whether conclusive statements
on the size of the pseudospin-symmetry breaking part of the 
proton-nucleus scattering amplitude can be made from polarization
and cross section data alone. Based on the exact relations
derived in Ref. \cite{Leeb00}, we formulate constraints on the 
polarization $P(\theta )$ and spin rotation $Q(\theta )$ in terms of
the aforementioned ratio $R_{\rm ps}$. It turns out that the 
polarization data provide a lower bound for $|R_{\rm ps}(\theta )|$. 
In addition, we show that a  lower bound of the absolute value 
of the pseudospin-symmetry breaking part of the proton-nucleus 
scattering amplitude can be extracted from
polarization and cross section data and thus a systematic study 
over a range of nuclei and energies can be made. 

In Sec. II we briefly outline the  scattering formalism in terms of 
the pseudospin-independent and pseudospin-dependent scattering
amplitudes. Based on  exact relations for the observables, we
formulate constraints on the pseudospin-symmetry breaking scattering
amplitude which require only the knowledge of the cross section and  
polarization. In Sec. III we present a systematic study of elastic 
proton scattering data concerning the size of the pseudospin 
symmetry breaking term.
In Sect. IV we discuss the role played by the Coulomb 
potential in the pseudospin symmetry breaking. Finally, we summarize
 our concluding remarks in Sec. V.  
\section{Derivation of Constraints}
\label{sec:II}
For the derivation of the constraints we must briefly recall the
formalism for the elastic scattering of a nucleon on a spin zero
target. The scattering amplitude $f(k,\theta )$, according to the standard 
notation of the literature (see, for example, Ref. \cite{Feshbach92}),  
is given by
\begin{equation}
     f(k,\theta ) = A(k,\theta ) + 
    B(k,\theta ) \mbox{\boldmath $\sigma $}{\cdot {\bf \hat n}}\, ,
\label{f}
\end{equation}
where $k$ is the momentum of the nucleon, \mbox{\boldmath $\sigma $} 
is the vector formed by the Pauli matrices, ${\bf \hat n}$ is the 
unit vector perpendicular to the scattering plane, and $\theta $ is 
the scattering angle. The complex-valued functions $A(k,\theta )$ 
and $B(k,\theta )$ are the spin-independent and spin-dependent 
parts of the scattering amplitude which are not fully accessible to
experiment. 

As shown by Ginocchio \cite{Ginocchio99b},
one can determine from the standard representation of the scattering 
amplitudes,  Eq. (\ref{f}), the pseudospin-independent 
${\widetilde A}$ and  pseudospin-dependent ${\widetilde B}$  
scattering amplitudes via a unitary transformation
\begin{equation}
\left( \begin{array}{c}
           \tilde A \\ \tilde B \end{array} \right) =
           \left( \begin{array}{cc} 
           \cos (\theta ) & \mbox{i}\sin (\theta ) \\
           \mbox{i}\sin (\theta ) & \cos (\theta ) 
\end{array} \right) \left( \begin{array}{c}
A \\ B \end{array} \right) \, .
\label{trafo}
\end{equation}
Partial wave expansions of $\tilde A$ and $\tilde B$ must be performed 
in terms of the pseudo-orbital angular momentum,
\begin{eqnarray}
\tilde \ell & = & \ell + 1\, , \mbox{ for } j = \ell + 1/2 =
                  \tilde \ell - 1/2\, ,
\nonumber \\
\tilde \ell & = & \ell - 1\, , \mbox{ for } j = \ell - 1/2 =
                  \tilde \ell +1/2\, ,
\end{eqnarray}
using the partial-wave S-matrix elements, $\tilde S_{\tilde \ell ,j}$,
defined for pseudo-orbital angular momentum
\begin{equation}
      \tilde S_{\tilde \ell,j=\tilde \ell -1/2} = 
       S_{\tilde \ell -1,j=\tilde \ell -1/2}\, , \quad  
       \tilde S_{\tilde \ell,j=\tilde \ell +1/2} = 
       S_{\tilde \ell +1,j=\tilde \ell +1/2}\, .
\end{equation}
For details of this transformation we refer to the original work
of Ginocchio \cite{Ginocchio99b}.

The observables in nucleon-nucleus scattering are usually described 
in terms of the amplitudes $A(k,\theta )$ and $B(k,\theta )$. Because
of the unitary transformation, Eq. (\ref{trafo}), they can be equally
described  in terms of $\tilde A(k,\theta )$ and $\tilde B(k,\theta )$.
For the scattering by a spinless target the observables are the 
differential cross section, 
\begin{equation}
       \frac{d\sigma }{d\Omega }(k,\theta ) = 
              |\tilde A(k,\theta )|^2 + |\tilde B(k,\theta )|^2\, ,
\label{ds}
\end{equation}
the polarization,
\begin{equation}
       P(k,\theta ) = 
              \frac{\tilde B(k,\theta )\tilde A^*(k,\theta )
               +\tilde B^*(k,\theta )\tilde A(k,\theta )}
              {|\tilde A(k,\theta )|^2 + |\tilde B(k,\theta )|^2} \, ,
\label{P}
\end{equation}
and the spin rotation function
\begin{widetext}
\begin{equation}
      Q(k,\theta ) = \frac{\sin (2\theta ) \left[ |\tilde A(k,\theta )|^2 
      - |\tilde B(k,\theta )|^2 \right] +\mbox{i} \cos (2\theta ) 
      \left[ \tilde B(k,\theta ) \tilde A^*(k,\theta )
        - \tilde B^*(k,\theta ) \tilde A(k,\theta )\right]}
      {|\tilde A(k,\theta )|^2 + |\tilde B(k,\theta )|^2}\, .
\label{Q}
\end{equation}
\end{widetext}
As can be shown, e.g. from Eqs. (\ref{P}) and (\ref{Q}),  $P^2+Q^2 \leq 1$.\\

The extraction of the full scattering amplitude (moduli and phases of 
$\tilde A(k,\theta )$ and $\tilde B(k,\theta )$) from measurements is
a very challenging task in quantum mechanics  intimately
related to the longstanding {\em phase problem} in diffraction
analyses (see, for instance, Refs. \cite{ph1,ph2,ph3}). 
Here, however, we are interested only in the formulation of constraints on 
$R_{\rm ps}(k,\theta )=\tilde B(k,\theta )/\tilde A(k,\theta )$ 
which yields a measure of the strength of the pseudospin-dependent 
part of the scattering. From Eqs. (\ref{ds}) to (\ref{Q}) it is
straightforward to write the observables in terms of the ratio 
$R_{\rm ps}$ (we suppress from now on the $k$-dependence)
\begin{equation}
      \frac{d\sigma(\theta) }{d\Omega} = 
                     |\tilde A(\theta)|^2 (1 + |R_{\rm ps}(\theta)|^2)\, ,
\label{sigmas}
\end{equation}
\begin{equation}
           P(\theta) = \frac{2\ Re (R_{\rm ps}(\theta))}
           {1 + |R_{\rm ps}(\theta)|^2}\, ,
\label{Ps}
\end{equation}
and 
\begin{equation}
           Q(\theta)= \frac{\left[ 1 - |R_{\rm ps}(\theta)
              |^2 \right]\sin (2\theta )
          - 2\ \mbox{i} \Im (R_{\rm ps}(\theta)) \cos (2\theta )}
         {1 + |R_{\rm ps}(\theta)|^2}\, .
\label{Qs}
\end{equation}
For pseudospin symmetry the ratio $R_{\rm ps}$ vanishes 
and consequently $P=0$ and $Q=\sin (2\theta )$ \cite{Bowlin88}, 
independent of $k$. One may also express  the ratio $R_{\rm ps}$ 
in terms of the polarization $P$ and the spin rotation $Q$ \cite{Leeb00}. 

\begin{figure}[t]
\centerline{\hbox{
        \epsfig{file=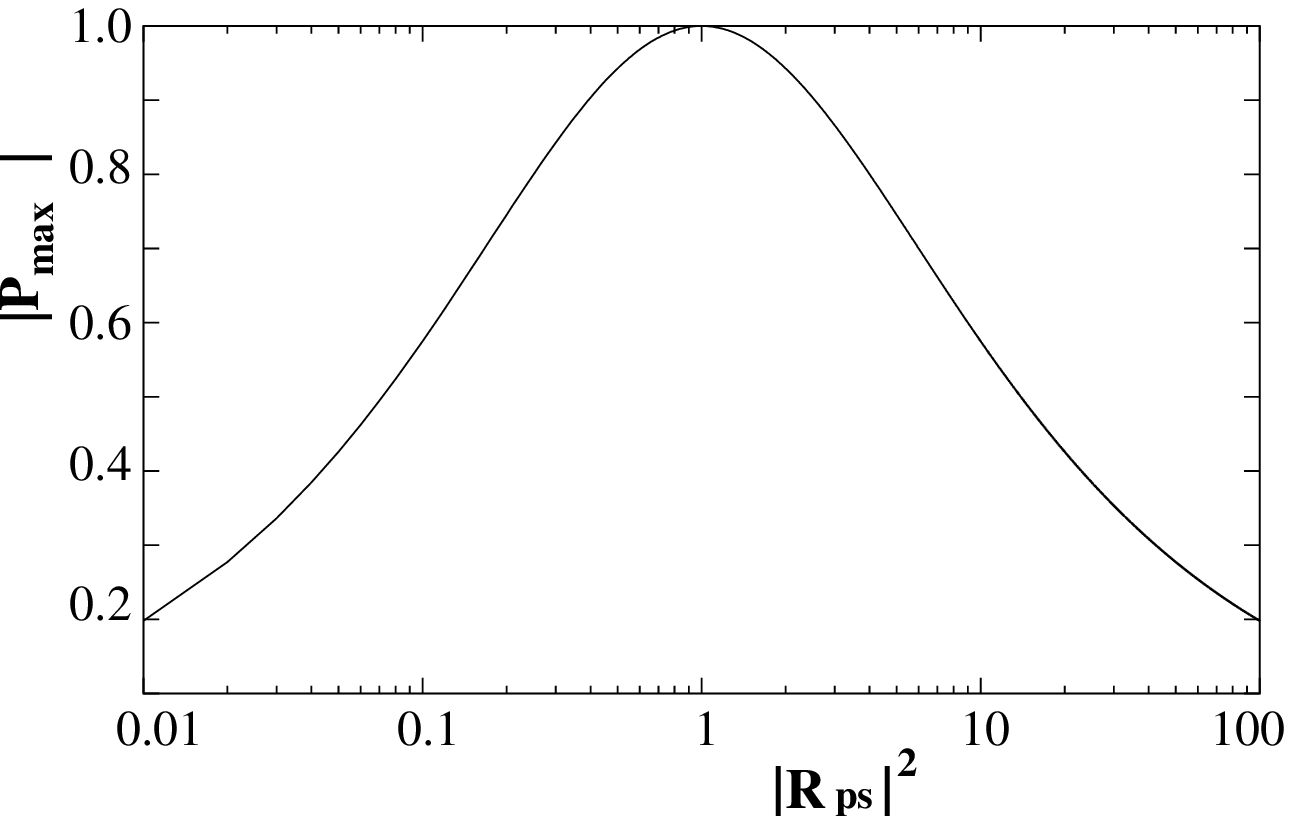,width=12cm}
}}
\bigskip
\caption{}
\label{fig1}
\end{figure}

From Eq. (\ref{Ps}) it is obvious that for a given value of $R_{\rm ps}$
the polarization is a constant, independent of the scattering angle 
$\theta $. 
Specifically, if we assume an upper admissible limit for the pseudospin 
symmetry breaking, i.e. if we assume  $|R_{\rm ps}|^2 \leq \Gamma $, 
we obtain the following bound for the polarization
\begin{equation}
      |P(\theta)| \leq \frac{2 \sqrt{\Gamma(\theta)}}{1+\Gamma(\theta)}\,.
\label{Pbound}
\end{equation}
The angle independence of the upper bound in $P(\theta )$ makes
it an ideal criterion to estimate from the polarization 
 the  pseudospin symmetry breaking term in nucleon-nucleus 
scattering. In Fig. \ref{fig1} we show the upper bound
$|P_{\rm max}|$ for a given $|R_{\rm ps}|^2$-value.

From this figure one can immediately extract the corresponding minimum 
and maximum values $\Gamma_{\rm min}$ and $\Gamma_{\rm max}$ for the ratio 
$|R_{\rm ps}|^2$ at each angle,
\begin{eqnarray}
       \sqrt{\Gamma_{\rm min}} & = & \frac{1}{|P|} [1-\sqrt{1-P^2}]
\label{rmin}
\, ,\\
       \sqrt{\Gamma_{\rm max}} & = & \frac{1}{|P|} [1+\sqrt{1-P^2}]
\, ,
\label{rmax}
\end{eqnarray}
Thus one can judge whether the pseudospin-symmetry breaking scattering
amplitude yields an important contribution in a certain angular range 
or not.

\begin{figure}[t]
\centerline{\hbox{
       \epsfig{file=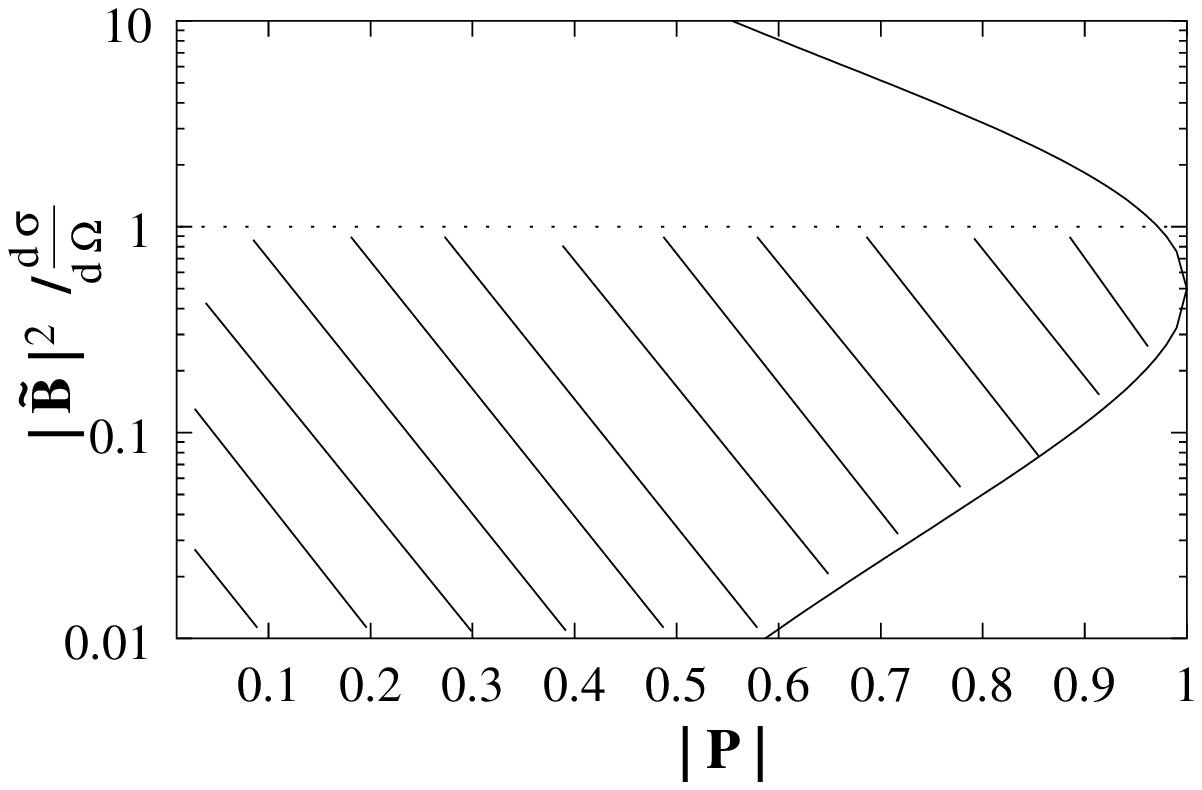,height=12cm,angle=270}
}}
\bigskip
\caption{}
\label{fig2}
\end{figure}

The ratio $R_{\rm ps}$ is not perhaps the best choice for an overall judgment
of the pseudospin symmetry breaking term, because it will exhibit
rather high values, when the pseudospin independent amplitude 
$\tilde A$ goes through minima in the angular range. Therefore,
for a more general consideration, an estimate of the absolute value 
of the pseudospin dependent scattering amplitude $\tilde B$ should 
accompany the analysis because it directly refers to the size of the 
contribution of the pseudospin dependent part. This amplitude can
be expressed in terms of the differential cross section (\ref{sigmas})
and the ratio $|R_{\rm ps}|^2$,
\begin{equation}
        |\tilde B(\theta)|^2 = \frac{|R_{\rm ps}(\theta)|^2}
    {1+|R_{\rm ps}(\theta)|^2} \ \frac{{\rm d}\sigma(\theta) }
       {{\rm d}\Omega }\, .
\label{Bexa}
\end{equation}
It was shown in Ref. \cite{Leeb00}, that $R_{\rm ps}$ can be 
fully determined from
experiment if the polarization and the spin-rotation function are measured. 
Since, however, in most cases the spin-rotation data are not available,
we must  look for an estimate of $|\tilde B|$ using 
differential cross section and polarization data alone. From Eq. 
(\ref{Bexa}) we  obtain 
\begin{equation}
      \frac{\Gamma_{\rm min}}{1+\Gamma_{\rm max}}\, 
       \frac{d\sigma }{d\Omega } \leq |\tilde B|^2 \leq 
     \frac{\Gamma_{\rm max}}{1+\Gamma_{\rm min}}\,
      \frac{d\sigma }{d\Omega }\, ,
\label{Best}
\end{equation}
where we have used the admissible range of $|R_{\rm ps}|$ determined
from the polarization via Eq. (\ref{rmin}) and (\ref{rmax}).
The ratio $|\tilde B|^2/\frac{d\sigma}{d\Omega}$, 
given by Eq. (\ref{Best}), as a function of the polarization $P$
is plotted in Fig. \ref{fig2}.
%
%
As can be seen from this figure, the boundaries  are 
useful when $P$ is in the vicinity of one.
Specifically, for $|P|\leq 0.972$ the upper bound is better 
estimated by $d \sigma $/$d \Omega$.
\section{Analysis of experimental data}
\label{sec:III}
\begin{figure}[t]
\centerline{\hbox{
        \epsfig{file=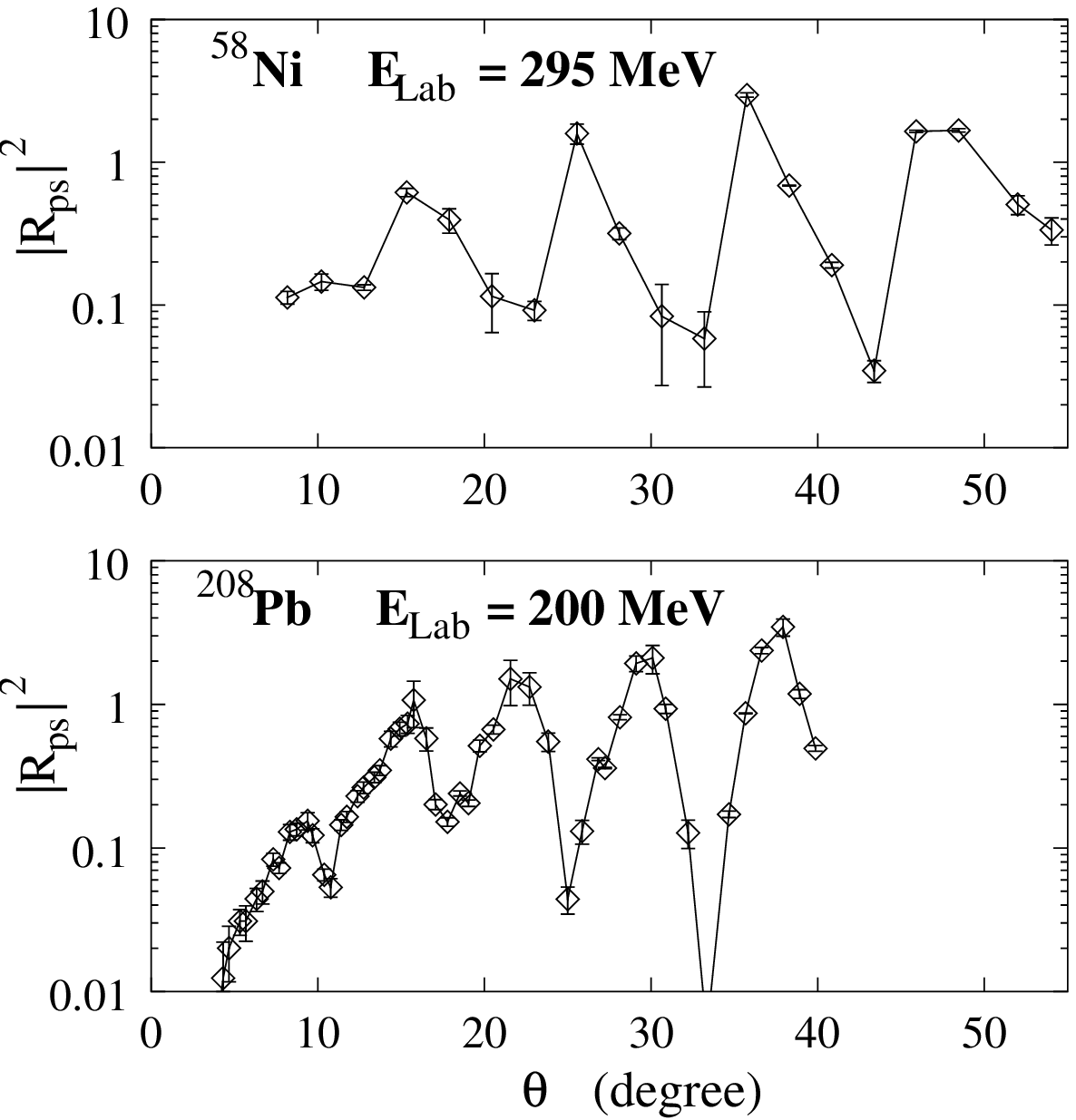,width=11cm}
}}
\bigskip
\caption{}
\label{fig3}
\end{figure}

The exact relation for $R_{\rm ps}$, derived in Ref. \cite{Leeb00},
can be directly applied to elastic proton-nucleus scattering data 
where measurements of the spin-rotation function $Q(\theta )$, 
 analyzing power $P(\theta )$, and  differential cross section 
${\rm d}\sigma(\theta )/{\rm d}\Omega$ are available. As
already mentioned, due to the difficulty in measuring 
the spin-rotation function, such measurements are scarce. 
Complete sets of measurements, however, are available e.g. 
for $^{58}$Ni at 295\,MeV \cite{Sakaguti98,RCNP}
and for $^{208}$Pb at 200\,MeV  \cite{Amos,Ji87}, and  800\,MeV
\cite{Blanpied78,Hoffmann78,Fergerson86}. The latter data have already 
been analyzed with respect to $|R_{\rm ps}|$ in Ref. \cite{Leeb00} and
no significant violation of the pseudospin symmetry at this energy 
was found. 
A similar analysis for the two data sets at  lower energies
is shown in Fig. \ref{fig3}. It is clear that at these 
energies the modulus of the ratio $R_{\rm ps}$ exhibits values which 
indicate a significant contribution of the pseudospin-dependent 
scattering amplitude within the range of the  measured angles. This 
finding of a stronger violation of the pseudospin-symmetry at lower 
energies is in qualitative agreement with the estimate of Bowlin 
{\em et al.} \cite{Bowlin88}.
%

\begin{figure}[t]
\centerline{\hbox{
        \epsfig{file=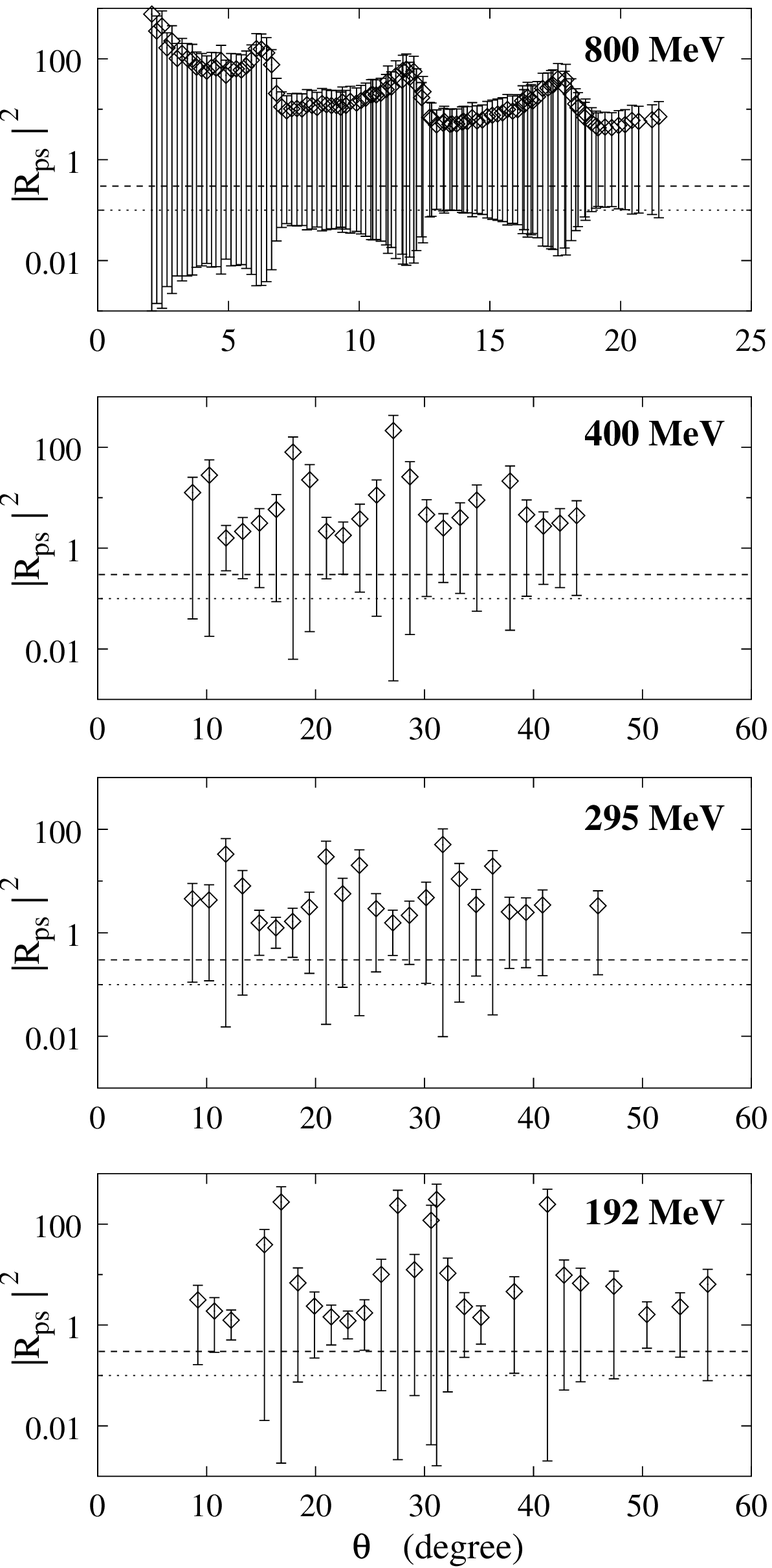,height=22cm}
}}
\bigskip
\caption{}
\label{fig4}
\end{figure}

In order to investigate further this finding, we studied the energy
dependence of the ratio $R_{\rm ps}$ by considering experimental 
p\,-$^{58}$Ni analyzing powers at  $E_{\rm Lab}= 192$\,MeV, 
295\,MeV, 400\,MeV \cite{Sakaguti98}, and 800\,MeV 
\cite{Hoffmann78}. Since spin-rotation data are not available for 
all data sets, we have applied the estimate 
$\Gamma_{\rm min} \leq |R_{\rm ps}|^2\leq \Gamma_{\rm max}$,
given in Eqs. (\ref{rmin}) and (\ref{rmax}), 
and the results obtained  are displayed in Fig.~\ref{fig4}. 
At low energies the estimate indicates again significant 
contributions to  pseudospin-symmetry violation  stemming from 
the pseudospin-dependent scattering amplitude at specific angles,
 in contrast to the small  violation 
observed at $E_{\rm Lab}=800$\,MeV. This 
systematics for the  p\,-$^{58}$Ni scattering confirms once more the
increased effect of pseudospin-symmetry violating contributions
at lower energies.

\begin{figure}[h]
\centerline{\hbox{
        \epsfig{file=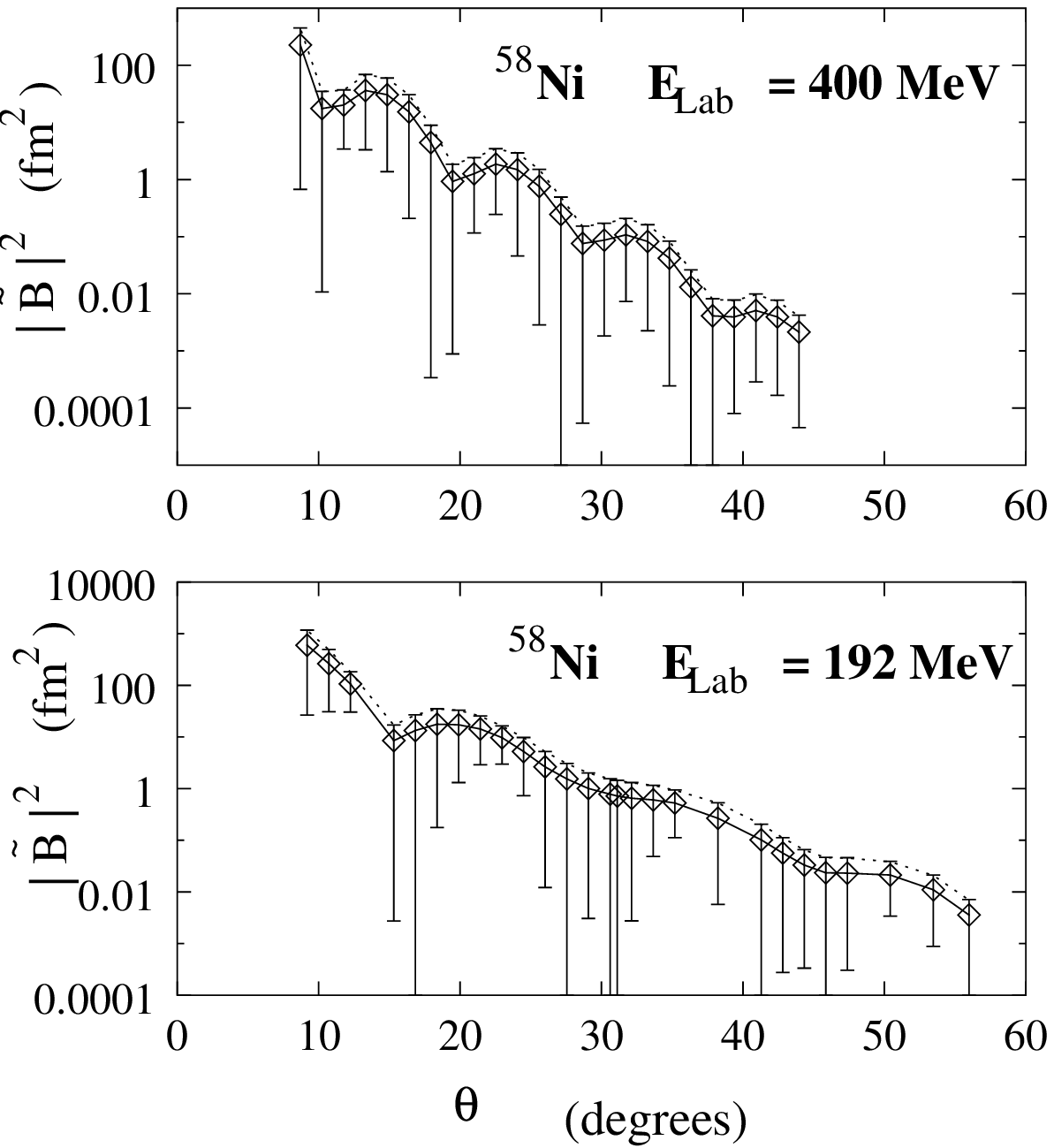,width=11cm}
}}
\bigskip
\caption{}
\label{fig5}
\end{figure}

As pointed out in section II, we consider also the absolute value of 
$\tilde B$ gives
a more direct measure for the size of the pseudospin-symmetry violation. 
For a complete data set, including $d\sigma /d\Omega(\theta ), 
 P(\theta )$, and $Q(\theta )$, the amplitude $|\tilde B|^2$ can be 
evaluated from  Eq. (\ref{Bexa}). In Fig. \ref{fig5} estimates for 
$|\tilde B|^2$ for p\,-$^{58}$Ni scattering at $E_{\rm Lab}=192$\,MeV 
and 400\,MeV from polarization and cross section data are given which
demonstrate the feasibility of the procedure by means of 
Eq. (\ref{Best}). To get a feeling about the relative size 
of $\tilde B$, the cross section 
data are also  shown for the purpose of comparison. In all cases 
considered, the admissible values for $|\tilde B|^2$ are of the same 
order as those of the cross sections. This  indirectly corroborates the 
significant pseudospin-symmetry violation in low energy proton-nucleus 
scattering. 
%
%

\section{The role of Coulomb potential}
The observation of weakly broken pseudospin-symmetry in proton and 
neutron shell orbit states is well established. As already mentioned
above, it is related to a symmetry of the Dirac Hamiltonian with 
$V_{\rm V}=-V_{\rm S}$, which is almost satisfied in relativistic mean field 
calculations. Such studies \cite{Ginocchio98a} yield a small 
pseudospin-symmetry breaking term which is necessary to explain the 
nuclear spectra \cite{Ginocchio97}. They lead to a splitting of 
quasi-degenerated states with a given $\tilde \ell $. The experimentally 
observed splittings are smaller than the theoretical ones, thus 
indicating that the actual pseudospin-symmetry breaking contributions 
are even weaker than expected from theory \cite{Alberto02}.
The role played by the Coulomb potential with regard to
pseudospin-symmetry breaking has recently been addressed by
Lisboa and Malheiro \cite{Lisboa02}. They found only weak
pseudospin-symmetry violation because significant cancellations
between nuclear and Coulombic terms occur.

At a first glance one would not expect drastic changes when going above 
threshold to the scattering region. However, the long range 
nature of the Coulomb interaction requires a special treatment. Albeit 
it is included in all scattering calculations, its role with regard to the 
pseudospin-symmetry has not been considered so far. In what follows we 
shall report some considerations on the Coulomb interaction in 
proton-nucleus scattering and its consequences with regard to 
pseudospin-symmetry.

\begin{figure}[h]
\centerline{\hbox{
        \epsfig{file=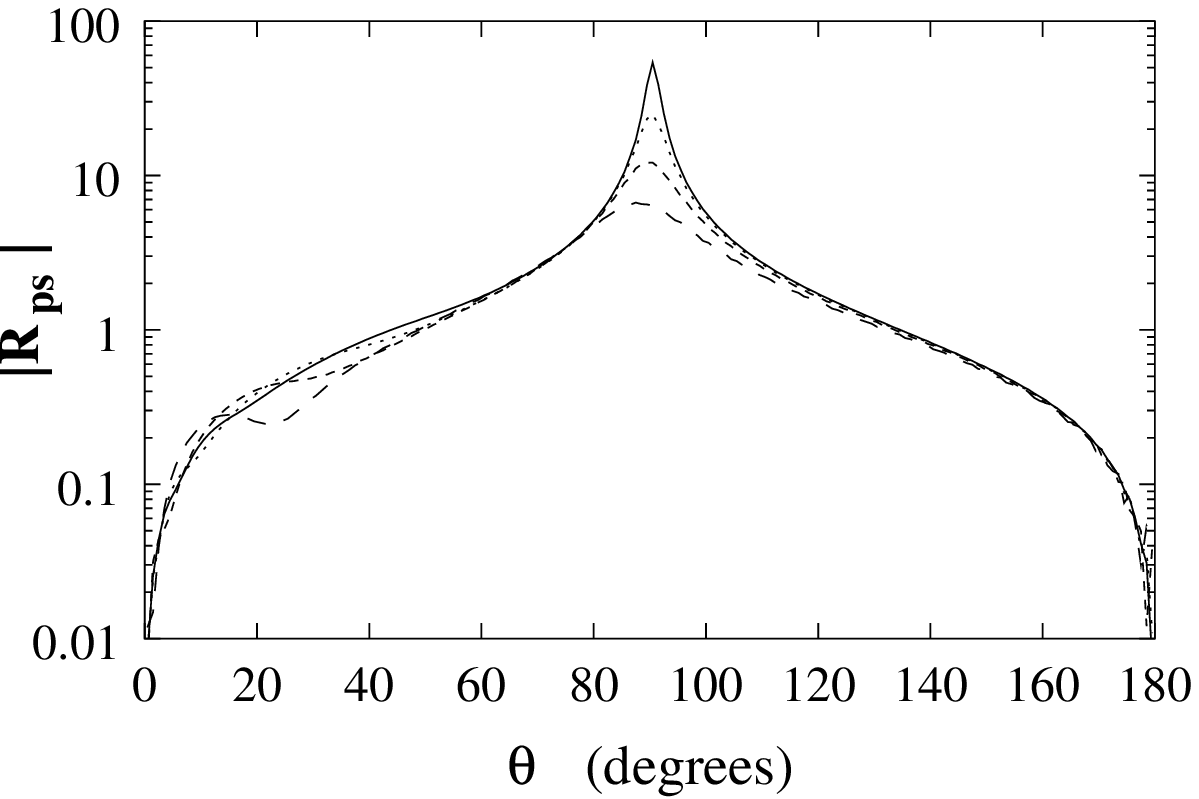,width=11cm}
}}
\bigskip
\caption{}
\label{fig6}
\end{figure}

First we consider the pure Coulomb scattering problem within the Dirac
equation for which closed-form expressions for the phase shifts
$\sigma_\ell^{({\rm C})}$ are known \cite{Gordon28},
\begin{equation}
       \exp(2i\sigma_\ell^\pm)=\frac{\gamma -i \eta}
           {\lambda -i{\bf \bar \eta}}
        \frac{\Gamma (\gamma +1+i\eta )}{\Gamma (\gamma +1-i\eta )}
       \,\exp(i\pi (\ell -\gamma )
\end{equation}
with
\begin{equation}
         \gamma =\sqrt{\lambda^2-Z^2\alpha_f^2}\,  \quad
         \eta =Z\alpha_f \frac{E}{\hbar k c}, \quad
        {\bf \bar \eta} =Z\alpha_f \frac{mc^2}{\hbar k c}
\end{equation}
and
\begin{equation}
\lambda = \left\{ \begin{array}{lcl} -(\ell +1) & \mbox{ for } & 
j=\ell +\smfrac{1}{2}\\ \ell & \mbox{ for } & j=\ell -\smfrac{1}{2}
\end{array} \right. \, .
\end{equation}
Here, $Z$ is the charge of the nucleus, $\ell $ the orbital angular 
momentum quantum number, $E$ the energy, and $k$ the wave number of the 
proton, respectively. The upper index $\pm $ refers to the angular
momentum quantum number $j=\ell\pm\smfrac{1}{2}$. With these phase 
shifts the Coulomb scattering amplitudes $A_{\rm C}(\theta )$ and 
$B_{\rm C}(\theta )$ can be evaluated. The corresponding amplitudes in
pseudospin representation $\tilde A_{\rm C}(\theta )$ 
and $\tilde B_{\rm C}(\theta )$
are obtained via Eq. (\ref{trafo}) and yield the ratio 
$R_{\rm ps}^{\rm C}=\tilde B_{\rm C}(\theta )/\tilde A_{\rm C}(\theta )$ 
which is a measure for the breaking of pseudospin-symmetry by the 
Coulomb interaction. In Fig. \ref{fig6} we show, as an example,
this ratio for proton-Pb scattering at different energies.
%
%
%
It is seen that pseudospin breaking due to the Coulomb potential is 
largest at 90 degrees and that it decreases with energy. 

A satisfactory description of proton-nucleus scattering is usually 
obtained within the Dirac phenomenology using a scalar potential 
$V_{\rm S}(r)$ and a fourth component of a vector potential 
$V_{\rm V}(r)$ which is composed of a nuclear part $V_{\rm V}^N(r)$ 
and the Coulomb potential $V_{\rm C}(r)$.
Both, $V_{\rm S}(r)$ and $V_{\rm V}^N(r)$ are complex potentials
 and are frequently taken to be of 
Woods-Saxon shape (see e.g. \cite{Geramb85}). 
The associated scattering amplitudes consist of three contributions,
\begin{eqnarray}
    A(\theta ) &=&  A_{\rm N}(\theta ) + A_{\rm C}(\theta ) 
                + A_{\rm I}(\theta )\, ,\\
    B(\theta ) &=&  B_{\rm N}(\theta ) + B_{\rm C}(\theta ) 
               + B_{\rm I}(\theta )\, .
\end{eqnarray}
Here, the indices N, C, and I  denote the nuclear, the Coulomb, and
the interference contributions, respectively. The latter takes into
account non-linear modifications of the scattering amplitudes due
to the superposition of the interactions $V_{\rm N}+V_{\rm C}$. Via Eq. 
(\ref{trafo}) one obtains also the corresponding pseudospin 
representations of the 
amplitudes $\tilde A(\theta )$ and $\tilde B(\theta )$. Due to
the short range nature of the nuclear interaction, the nuclear parts of
the scattering amplitudes decrease more rapidly than the Coulomb
parts with increasing momentum transfer. Therefore, the ratio
$R_{\rm ps}(\theta )$ at increasing scattering angle will be dominated 
by the Coulomb contribution. 

In order to demonstrate this characteristic behavior of the 
scattering amplitudes we study a schematic example of proton-$^{208}$Pb
scattering assuming $V_{\rm S}(r)$ and $V_{\rm V}^{\rm N}(r)$ to 
be real and of Woods-Saxon shape
\begin{eqnarray}
     V_{\rm S}(r)  &=& W_0 \left[1+\exp (\frac{r-R_0}{a})\right]^{-1}\, ,\\
     V_{\rm V}^{\rm N}(r)&=& V_0 \left[1+\exp (\frac{r-R_0}{a})\right]^{-1}\, ,
\end{eqnarray}
with the half density radius $R_0=1.25$\,fm\,A$^{1/3}$ and diffuseness 
$a=0.6$\,fm. The Coulomb potential $V_{\rm C}(r)$ is that of a homogeneously
charged sphere with radius $R_{\rm C}=1.25$\,A$^{1/3}$. In the first example
we consider the nuclear strengths $V_0 = -W_0 = 300$\,MeV which 
implies exact pseudospin-symmetry in the nuclear part. The corresponding
scattering amplitudes have been evaluated for several energies between
$E_{\rm Lab}=200$\,MeV and $800$\,MeV.

\begin{figure}[h]
\centerline{\hbox{
        \epsfig{file=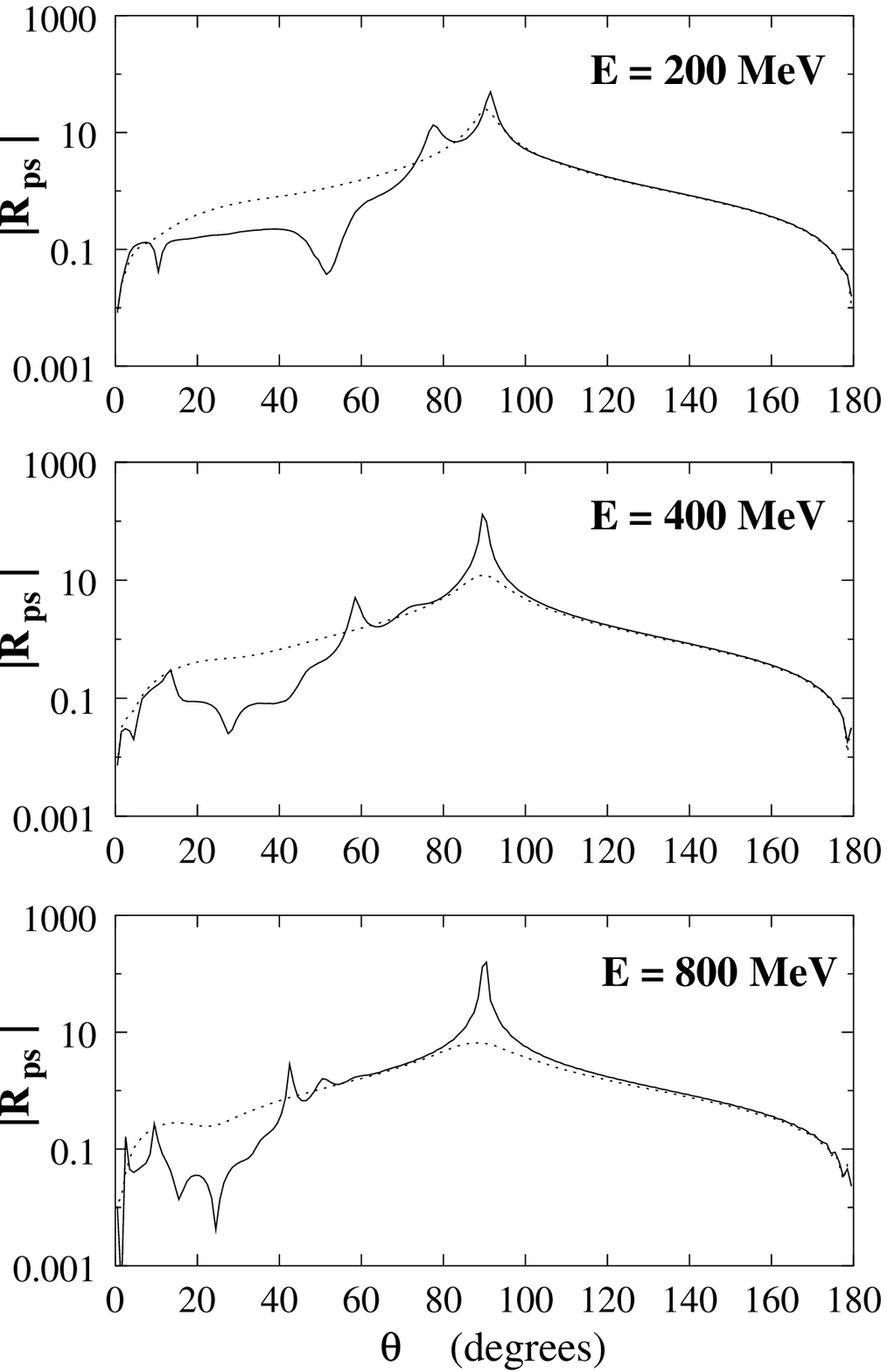,width=11cm}
}}
\bigskip
\caption{}
\label{fig7}
\end{figure}

%
%
 In Fig. \ref{fig7} the angular 
dependence of the absolute values of the ratio $R_{\rm ps}(\theta )$ is 
compared with that of the pure Coulomb interaction.  
This comparison clearly indicates that the ratio $R_{\rm ps}$ approaches
at backward angles the values of the pure Coulomb interaction. Hence,
at all energies considered, there exists an angular range with 
$|R_{\rm ps}|$-values not compatible with pseudospin-symmetry. The
non-vanishing $|R_{\rm ps}|$-values are solely caused by the presence of
the Coulomb potential since the nuclear part alone satisfies, because
of $V_{\rm V}^{\rm N}(r)=-V_{\rm S}(r)$, an exact pseudospin-symmetry. 

\begin{figure}[h]
\centerline{\hbox{
        \epsfig{file=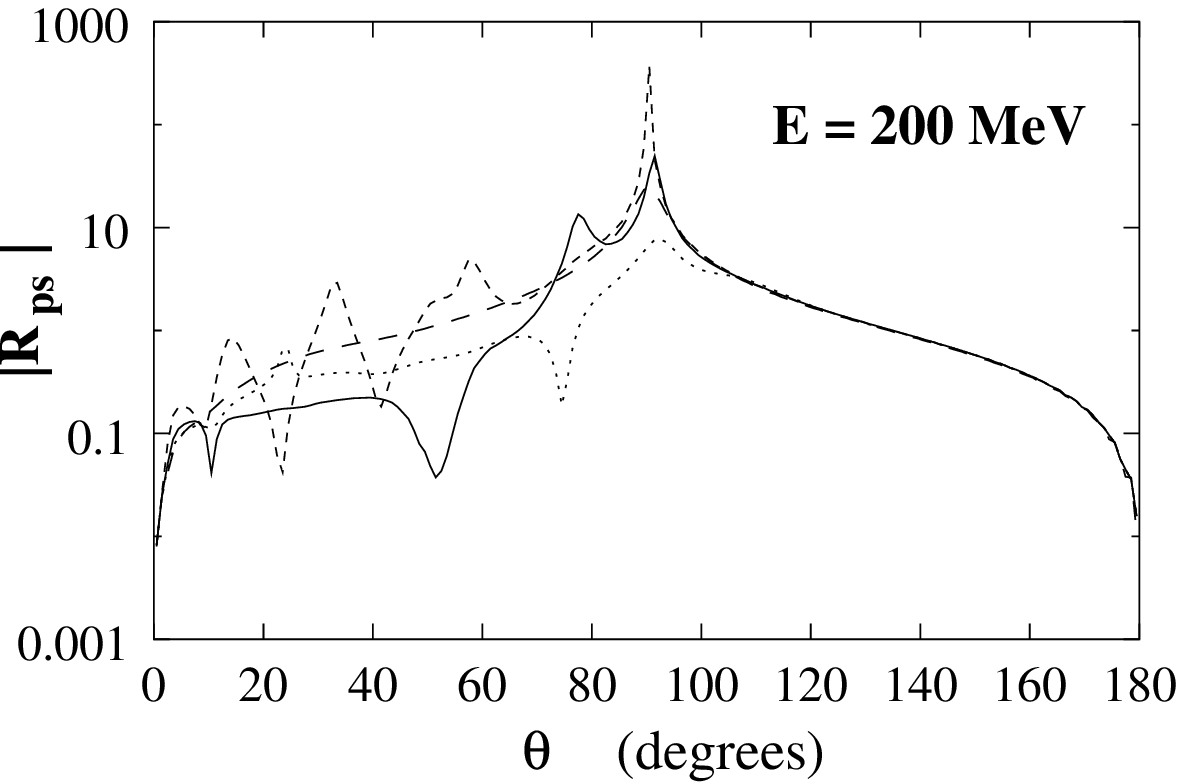,width=11cm}
}}
\bigskip
\caption{}
\label{fig8}
\end{figure}

Relativistic mean field calculations reveal a small but necessary
pseudospin-symmetry breaking nuclear part. In order to simulate this
effect we consider the same scattering system as before but with
strengths $V_0=-W_0+\Delta = 300$\, MeV and thus  
the pseudospin-symmetry is broken by the extra strength $\Delta $. 
In Fig. \ref{fig8} the moduli of the corresponding ratio 
$R_{\rm ps}(\theta )$ are shown for different $\Delta $-values 
at $E_{\rm Lab}=200$\,MeV. 

%
%
Qualitatively we observe the same behavior as in the case of $\Delta =0$.
There are again significant values of $|R_{\rm ps}|$ at backward angles which
are not in agreement with pseudospin-symmetry. It should be emphasized
that in all cases and at all angles considered the $|R_{\rm ps}|$-values 
associated with the nuclear part alone do not exceed $0.15$ thus 
indicating a small pseudospin-symmetry violating contribution.  

\begin{figure}[h]
\centerline{\hbox{
        \epsfig{file=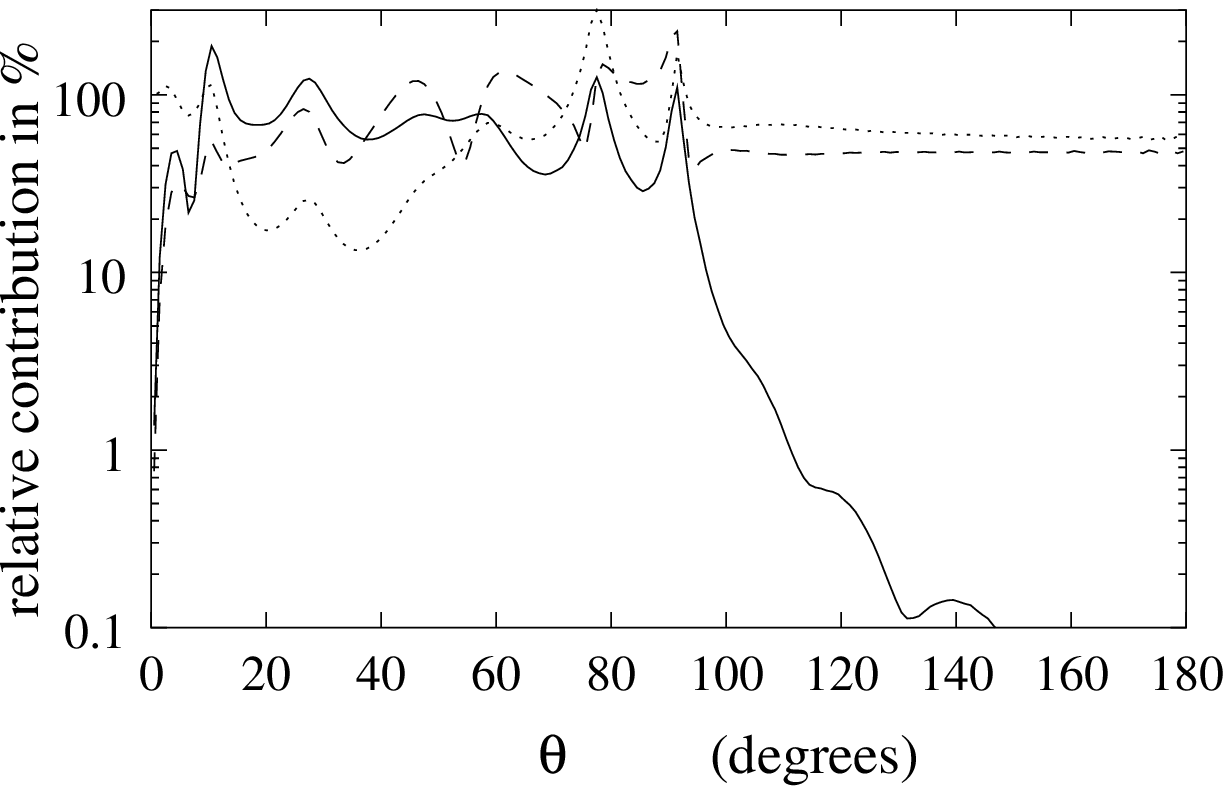,width=11cm}
}}
\bigskip
\caption{}
\label{fig9}
\end{figure}

Unfortunately it is not possible to extract the nuclear contribution
from the experimental proton-nucleus scattering data, the main reason
being  the non-linear relationship between potential and scattering 
amplitudes in the energy region considered. The importance of higher
order Born terms is best reflected in the importance of the amplitude
$|A_{\rm I}(\theta )|$. In Fig. \ref{fig9} the modulus of 
$A_{\rm I}(\theta )$ is compared with that of $A_{\rm N}(\theta )$ 
and $A_{\rm C}(\theta )$ which are of the same size. 
%
%
It is clear that it is not possible at  present to separate the nuclear
term without further model assumptions.
\section{Summary}
We have derived boundaries for the ratio of the pseudospin-dependent
to the pseudospin-independent scattering amplitude requiring only 
the knowledge of polarization data. In addition, we also derived 
boundaries for the absolute size of the pseudospin dependent 
scattering amplitude. These boundaries are based on the differential cross 
section and polarization data alone and, thus, one  can avoid  measurements 
of the spin-rotation function as required by the methods of Refs. 
\cite{Ginocchio99b,Leeb00}. Because of the difficulties in 
measuring $Q(\theta )$, these constraints could be very
useful and represent an improvement with regard to the previous 
situation. Their use together with the exact relationships
derived in \cite{Leeb00} allow us to assess the relevance of 
pseudospin symmetry in proton-nucleus scattering at various
energies. 
Furthermore, by considering p\,-$^{58}$Ni and 
p\,-$^{208}$Pb scattering data the mass number dependence 
is also shown up. 

The results for the ratio $|R_{\rm ps}|^2$ exhibit a systematic 
decrease with increasing energy.
At lower proton-nucleus scattering energies, up to  about $400$\,MeV, 
the results obtained indicate that there is a significant symmetry 
breaking term present which confirms the conjecture of 
Bowlin {\emph et al.} \cite{Bowlin88} 
based on analytical estimates.  At these energies and  at certain 
scattering angles the extracted $|R_{\rm ps}|^2$-values 
approach 1, implying that the pseudospin dependent and independent 
scattering amplitudes are of comparable size. 
It should be emphasized here that estimates of $|\tilde B|^2$ exhibit 
an angular dependence which is quite similar in form and size to 
that of  the differential cross section. 
In contrast, at 800\,MeV there is only a weak  violation of 
pseudospin symmetry  which  confirms the finding 
of \cite{Ginocchio99b,Leeb00}.

To investigate the origin of this behavior, we have studied the role 
played by the Coulomb interaction. Considering the Coulomb scattering
problem within the Dirac equation, leads to scattering amplitudes
$\tilde A(\theta )$ and $\tilde B(\theta )$ whose ratio $|R_{\rm ps}|$
shows a clear peak, with values greater than $1$ at $90$ degrees, which 
decreases with energy. Assuming typical  nuclear interactions 
of the  Dirac phenomenology, we have evaluated the ratio 
$R_{\rm ps}$ in the presence of the Coulomb interaction. The results 
clearly indicate that  the modulus of the ratio $|R_{\rm ps}|$, 
at all energies, exceeds $1$ and approach at backward 
angles that of the Coulomb problem.

Smaller $|R_{\rm ps}|$-values are found at small scattering angles 
which can be attributed  to highly non-linear superposition 
of nuclear and Coulomb effects.
One might conjecture that this phenomenon is of the same nature
to  the cancellation effects observed in the bound state regime 
\cite{Alberto02}. Anyway it is limited to a small and energy dependent 
angular region reflecting the finite range of the nuclear interaction. 
Because of the importance of nonlinearities between potential and 
scattering amplitude (higher order Born terms) it is impossible 
to separate from experimental data the nuclear components unambiguously 
without further model assumptions. In addition, we didn't find any 
characteristic behavior of $R_{\rm ps}(\theta )$ which would give a 
clear indication of a weakly pseudospin-symmetry violating nuclear term.   

In short, proton-nucleus scattering does not exhibit the features of
pseudospin-symmetry. The violation of the symmetry stems from the
long range Coulomb interaction and shows up in the values of the 
ratio $R_{\rm ps}$ at large scattering angles. The previous finding of
small values of the moduli of $R_{\rm ps}$ at $800$\,MeV proton-$^{208}$Pb
scattering, can be attributed to the limited angular range of 
the experimental data and cannot be considered as a clear sign 
for the relevance of pseudospin-symmetry in proton-nucleus scattering. 

\acknowledgments
The authors are indebted to Profs. K. Amos, R. Hofmann, L. Ray, and 
H. Sakaguchi for providing us the proton-nucleus scattering data in 
tabular form. One of us (H.L.) wants to thank the Department of 
Physics of the University of South Africa for the hospitality 
and  a financial support.

%
%
\vfill
\newpage

\section*{Figure Captions}

\bigskip
\noindent {\bf Fig. 1:} \quad
The upper limit of the polarization for a given 
pseudospin breaking ratio $|R_{\rm ps}|^2$.

\bigskip

\noindent {\bf Fig. 2:} \quad
The lower and upper bounds (solid lines) of the
ratio $|\tilde B|^2/\frac{d\sigma}{d\Omega}$ given by Eq. 
(\ref{Best}) as a function of the polarization $P$. In
addition the upper bound due to the cross section is shown
by the dotted line. The shaded area shows the admissible 
range.

\bigskip

\noindent {\bf Fig. 3:} \quad
Angle dependence of the ratio $|R_{\rm ps}|^2$ 
extracted from complete data sets.

\bigskip

\noindent {\bf Fig. 4:} \quad
The range of $|R_{\rm ps}|^2$ for p\,-$^{58}$Ni scattering 
extracted from polarization data at different energies $E_{\rm Lab}$. 
For a better estimate of the size, the values $|R_{\rm ps}|^2=0.3$ and 
$|R_{\rm ps}|^2=0.6$ are shown by a dotted and a dashed line, 
respectively.

\bigskip

\noindent {\bf Fig. 5:} \quad
The values $|\tilde B|^2$ extracted via Eq. (\ref{Best}) from
experimental proton-$^{58}$Ni polarization and scattering cross
section data at $E_{\rm Lab}=192$\,MeV and at $400$\,MeV. For 
comparison, the cross sections are also shown by a dotted line.

\bigskip

\noindent {\bf Fig. 6:} \quad
The modulus of the ratio $R_{\rm ps}(\theta )$ for pure Coulomb 
scattering of a proton by a Pb-nucleus at $E_{\rm Lab}=100$\,MeV 
(solid line), $E_{\rm Lab}=200$\,MeV (dotted line), 
$E_{\rm Lab}=400$\,MeV (dashed line), and $E_{\rm Lab}=800$\,MeV 
(long dashed line).
\bigskip

\noindent {\bf Fig. 7:} \quad
The modulus of the ratio $R_{ps}(\theta )$ evaluated with a schematic 
Dirac potential for proton-$^{208}$Pb scattering at several energies. 
The corresponding ratio obtained for pure Coulomb scattering is also
shown for comparison by dashed line. The nuclear part of potential 
satisfies pseudospin-symmetry. See text for more details.

\bigskip

\noindent {\bf Fig. 8:} \quad
The modulus of $R_{ps}(\theta )$ obtained from model calculations for
proton-$^{208}$Pb scattering at $E=200$\ MeV assuming different
strength $\Delta $ of the nuclear pseudospin-symmetry breaking term.
The results for $\Delta = 0$\ MeV (solid line), $\Delta =-50$\ MeV
(dotted line), and $\Delta = +50$\ MeV are shown. For comparison also
the absolute values of $R_{ps}$ for pure Coulomb scattering are 
given by long dashed line. 

\bigskip

\noindent {\bf Fig. 9:} \quad
The relative contributions of $\tilde A_N(\theta ), \tilde A_C(\theta )$,
and $\tilde A_I(\theta )$ to the scattering amplitude $\tilde A(\theta )$  
for proton-$^{208}$Pb scattering at $E=200$\ MeV. The values are
evaluated with the Dirac potentials given in Fig. 7. The quantities
$|\tilde A_N(\theta )/\tilde A(\theta )|$ (solid line),
$|\tilde A_C(\theta )/\tilde A(\theta )|$ (dotted line), and
$|\tilde A_I(\theta )/\tilde A(\theta )|$ (dashed line) are displayed.

\vfill

\newpage

\phantom{a}

\vskip5cm

\phantom{a}

\vfill

\newpage

\phantom{a}

\vskip5cm

\phantom{a}

\vfill

\newpage

\bigskip

\vfill

\newpage

\vfill

\newpage

\vfill

\newpage

\vfill

\newpage

\vfill

\newpage

\vfill

\newpage

\end{document}